\documentclass[11pt,twocolumn]{article}
\usepackage[utf8]{inputenc}
\usepackage[a4paper,margin=1in]{geometry}
\usepackage{graphicx}
\usepackage{ulem}

\usepackage[percent]{overpic}
\usepackage{color}
\usepackage{amsmath}

\title{Tailored optical potentials for Cs atoms above waveguides with focusing dielectric nano-antenna}
\author{Angeleene S. Ang$^{1,2\dagger}$, Alexander S. Shalin$^3$, Alina Karabchevsky$^{1,2*}$
\\
{\small $^1$ School of Electrical and Computer Engineering, Ben-Gurion University of the Negev, Israel}
\\
{\small $^2$ Center for quantum science and technology (BGU-QST), Ben-Gurion University of the Negev, Israel }
\\
{\small $^3$ ITMO University, 49 Kronversky Ave., St. Petersburg, Russia }
\\
{\small $*$ \texttt{alinak@bgu.ac.il} }
\\
{\small $\dagger$ \texttt{angeleene.ang@gmail.com} }
}

\begin{document}

\maketitle

\section*{Abstract}
Tuning the near-field using all-dielectric nano-antennae offers a promising approach for trapping atoms, which could enable strong single-atom/photon coupling.
Here we report the numerical study of an optical trapping of single Cs atom above a waveguide with silicon nano-antenna which produces a trapping potential for atoms in a chip-scale configuration.
Using counter-propagating incident fields, bichromatically detuned from the atomic cesium D-lines, we numerically investigate the dependence of the optical potential on the nano-antenna geometry.
{We tailor the near-field potential landscape by tuning the evanescent field of the waveguide using a toroidal nano-antenna, a configuration that enables trapping of ultracold Cs atoms.} 
Our research opens up a plethora of trapping atoms applications in a chip-scale manner from  quantum computing, to quantum sensing to list a few.

\vspace{2em}

{Controlled positioning of atoms and molecules with extreme precision is a fascinating achievement in the field of quantum optics and photonics.
Optical trapping was first shown by Ashkin in his seminal work (1970), where he trapped a dielectric microparticle (2018 Nobel Prize in Physics); in 1978, he suggested a three-dimensional trap for neutral atoms\cite{ashkinPRL1978}.}
Since then, the optical trapping of particles and, specifically, atoms have been extensively investigated \cite{grimmAIAMaOP2000,gustavsonPRL2001,kimNC2019}.
So far, the smallest optical atomic traps are realized with dimensions of half an optical wavelength.
However, this is not a fundamental limit.
Atomic trapping can be achieved at even smaller distances - in the sub-wavelength regime - by tuning near-fields in order to overcome the diffraction limit of far-fields.

In the case of confining a particle much smaller than the wavelength of the trapping light (Rayleigh regime), the trapping force is directly proportional to the intensity gradient of the field.
High-power lasers have conventionally been used to generate a strong intensity gradient, for the trapping and manipulation of particles.
To address the need for high laser power, optical trapping using plasmonic nanostructures or metamaterials\cite{stehleNP2011} have been proposed, as plasmonic systems allows localisation of electromagnetic radiation at the nanoscale\cite{karabchevskyLSA2016,galutinSR2017}.
{The plasmonic structures used in the aforementioned work are traditionally made of metals, which suffer from heating effects\cite{brongersmaS2010, ndukaifeS2016}, 
leading to Johnson noise, and limited storage time of atom-based systems\cite{gehmPRA1998, savardPRA1997}.}
Previous work has, nevertheless, experimentally demonstrated atomic traps utilizing plasmonic structures\cite{chenPRP2017,stehleNP2011}.
{On the other hand, photonic nanojets - narrow, high-intensity light beams emerging from the shadow side of dielectric micro/nanostructures - generated by spheres embedded on top of a substrate illuminated by a plane wave, have been proposed as possible optical atomic traps\cite{yannopapasOC2012,yannopapasJPCM2009}.}
By removing the metallic element, an all-dielectric set-up greatly reduces the noise, and it has been shown that such systems, e.g., based on optical fibers\cite{sorensenPRL2016, corzoPRL2016,vetschPRL2010,gobanPRL2012}, can provide electric fields sufficient for atom trapping.

\begin{figure}
\centering
      \begin{overpic}[width=0.65\linewidth, tics=10]{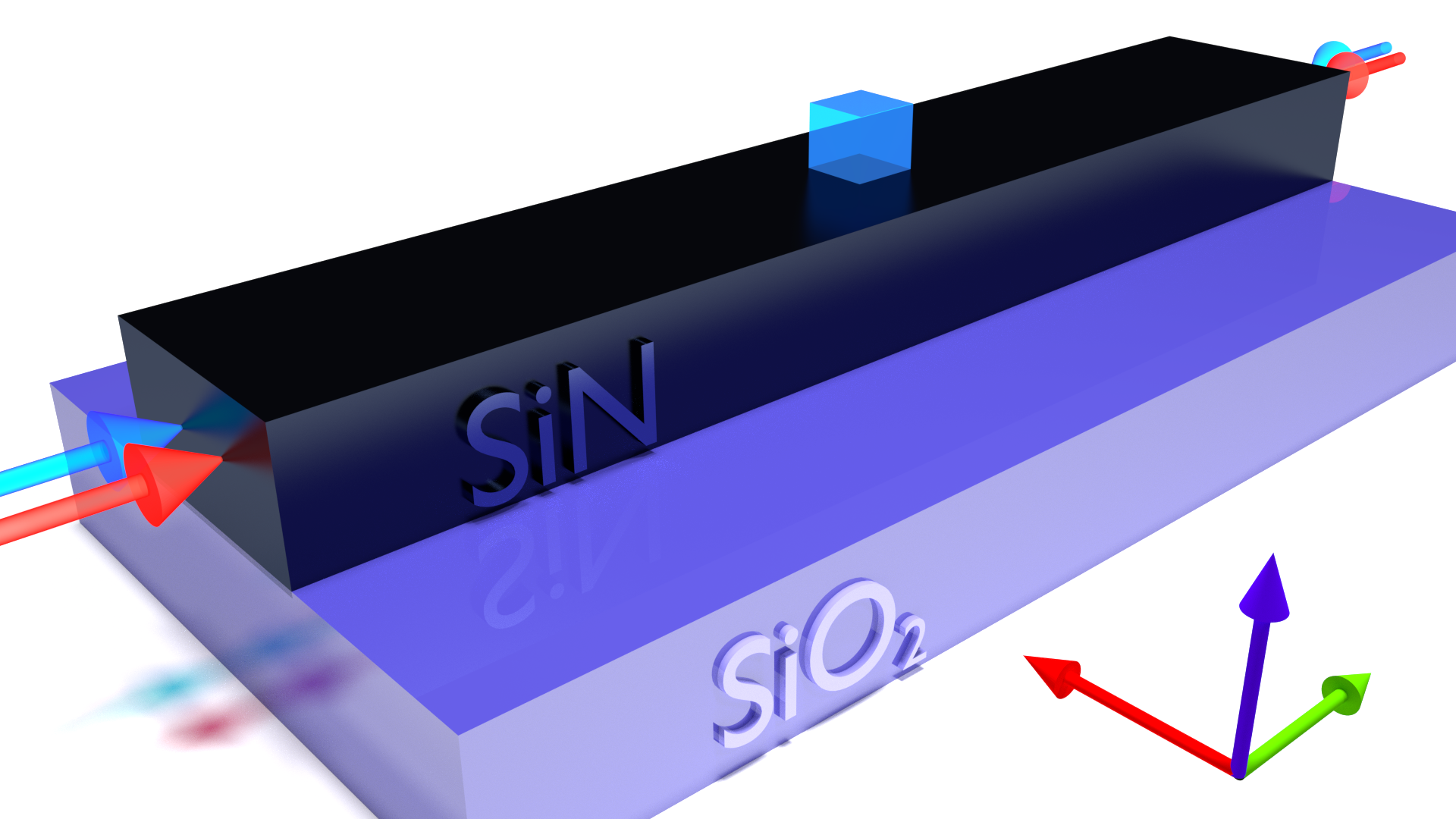}
         \put(-20,25){\rotatebox{18}{\large $\lambda_{blue}$}}
         \put(-15,10){\rotatebox{18}{\large $\lambda_{red}$}}
         \put(95,11){\color{green} \large $x$}
         \put(75,13){\color{red} \large $y$}
         \put(87,20){\color{blue} \large $z$}
      \end{overpic}
\caption{Illustration of the ridge waveguide, bi-chromatic incident light, and the silicon nano-antenna (not to scale). 
}
\label{fig:large_waveguide_and_inset}
\end{figure}
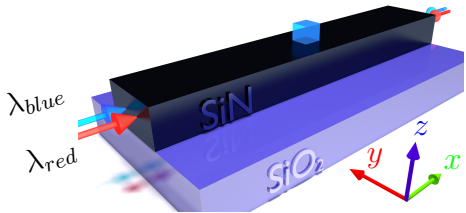

Hence, we propose utilizing an all-dielectric system - nano-antennae embedded on top of a ridge waveguide - as an alternative atomic trapping set-up to mitigate the heating issues while trapping atoms.
In this paper, we numerically investigate the dependence of the bichromatically constructed trapping potentials for cesium atoms on the nano-antenna shape.
Our system is illustrated in Fig.~(\ref{fig:large_waveguide_and_inset}):
we use counter-propagating incident bichromatic fields launched into both end-facets of the waveguide in order to generate a standing wave for trapping along the propagation direction; the silicon nano-antenna placed on the waveguide then localizes the light further along the other directions.

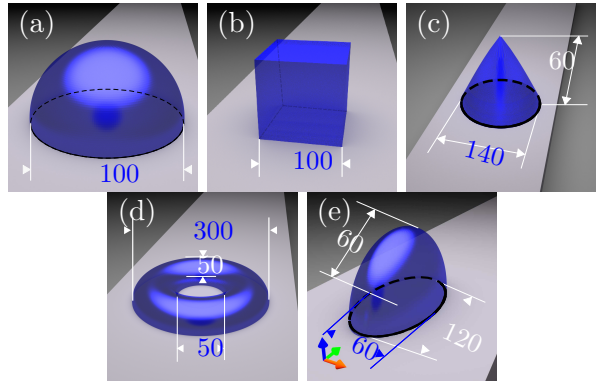
\begin{figure}[h]
\centering
\begin{overpic}[width=0.32\linewidth,tics=20]{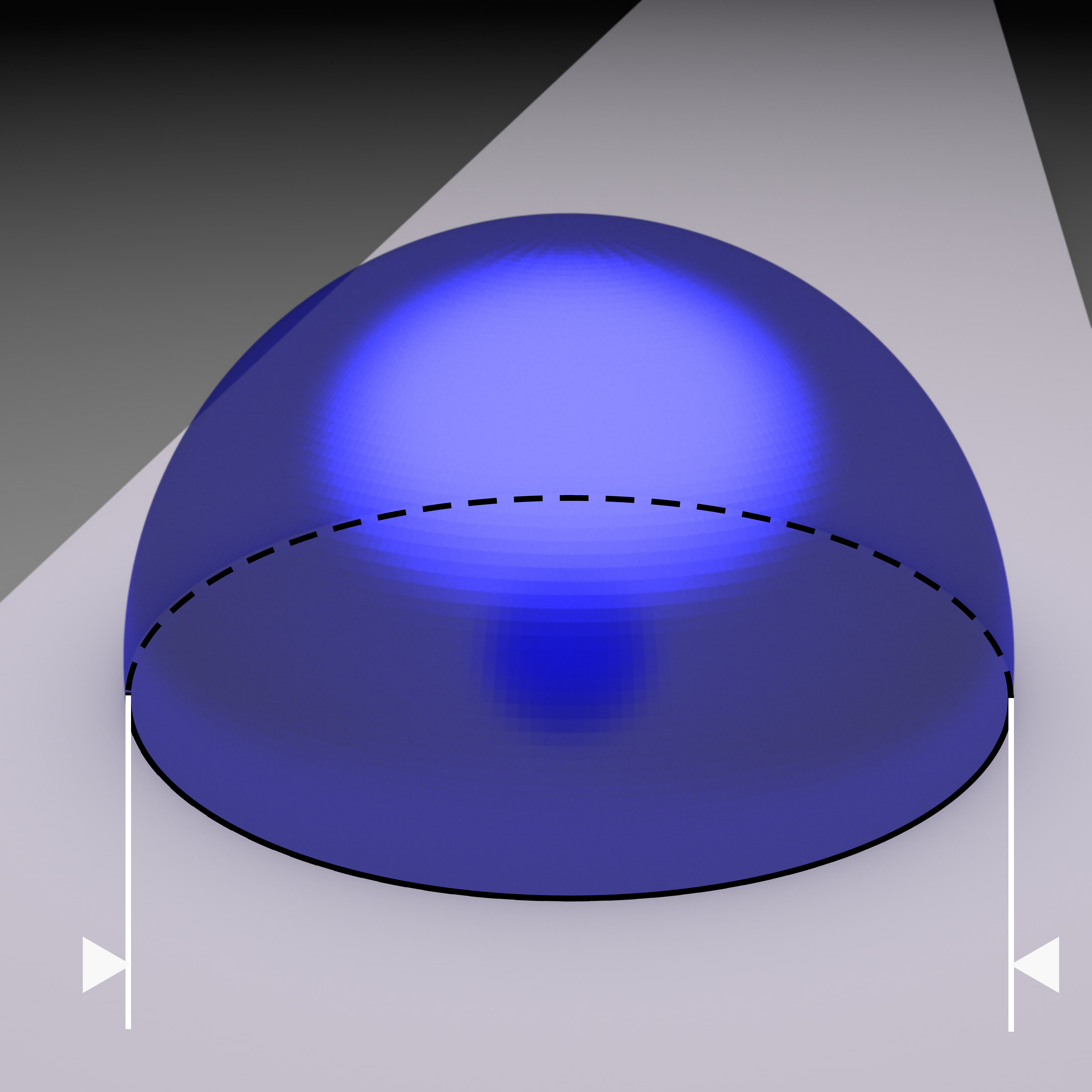}
\put(0,85){ \color{white} (a) }
  \put(43,5){\rotatebox{0}{ \small \color{blue} 100 }}
\end{overpic}
\begin{overpic}[width=0.32\linewidth,tics=20]{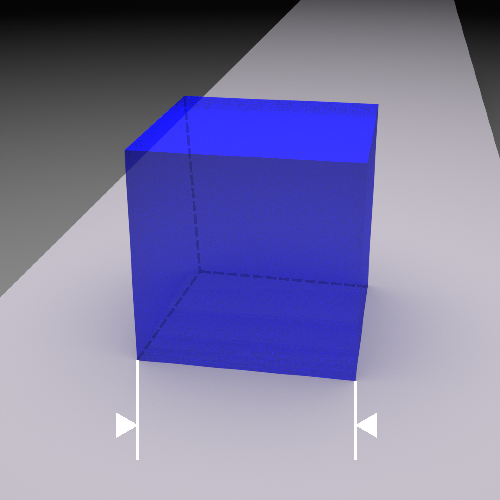}
\put(0,85){ \color{white} (b) }
  \put(40,12){\rotatebox{0}{ \small \color{blue} 100 }}
\end{overpic}
  \begin{overpic}[width=0.32\linewidth,tics=20]{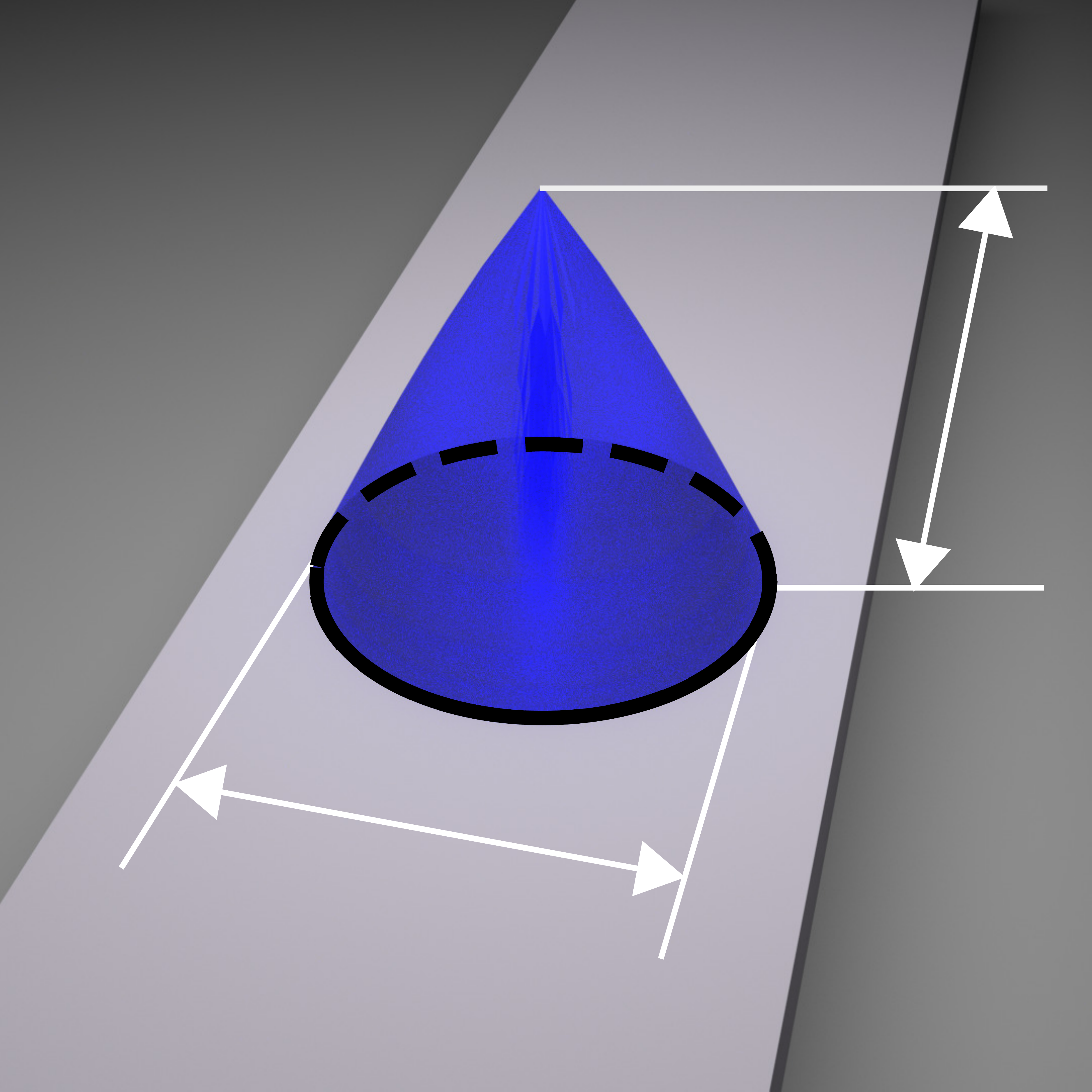}
  \put(0,85){ \color{white} (c) }
    \put(25,20){\rotatebox{-15}{ \small \color{blue} 140 }}
    \put(70,65){\rotatebox{0}{ \small \color{white} 60 }}
  \end{overpic}

    \begin{overpic}[width=0.32\linewidth,tics=20]{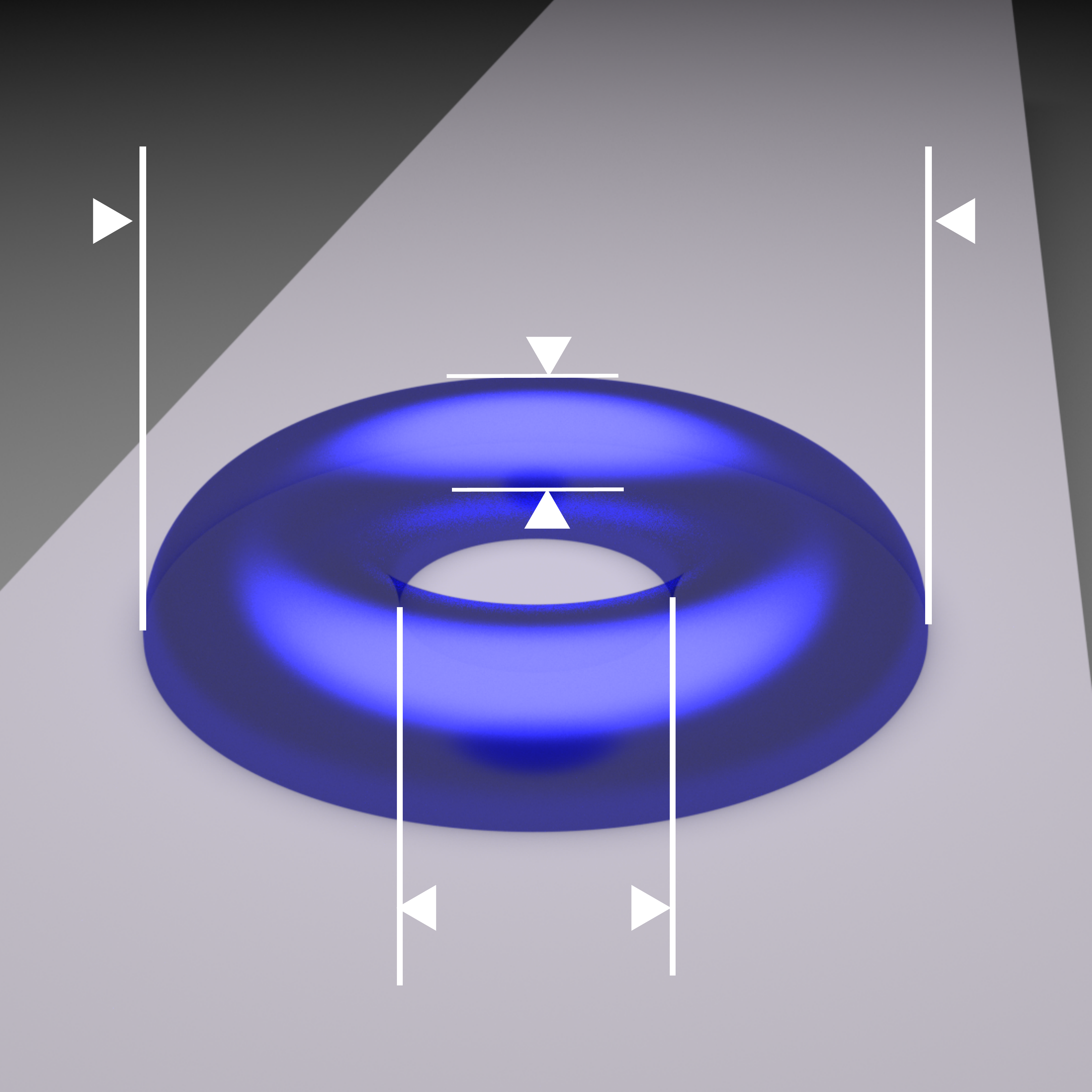}
    \put(0,85){ \color{white} (d) }
      \put(40,75){\rotatebox{0}{ \small \color{blue} 300 }}
      \put(42,13){\rotatebox{0}{ \small \color{blue} 50 }}
      \put(42,57){\rotatebox{0}{ \small \color{white} 50 }}
    \end{overpic}
       \begin{overpic}[width=0.32\linewidth,tics=10]{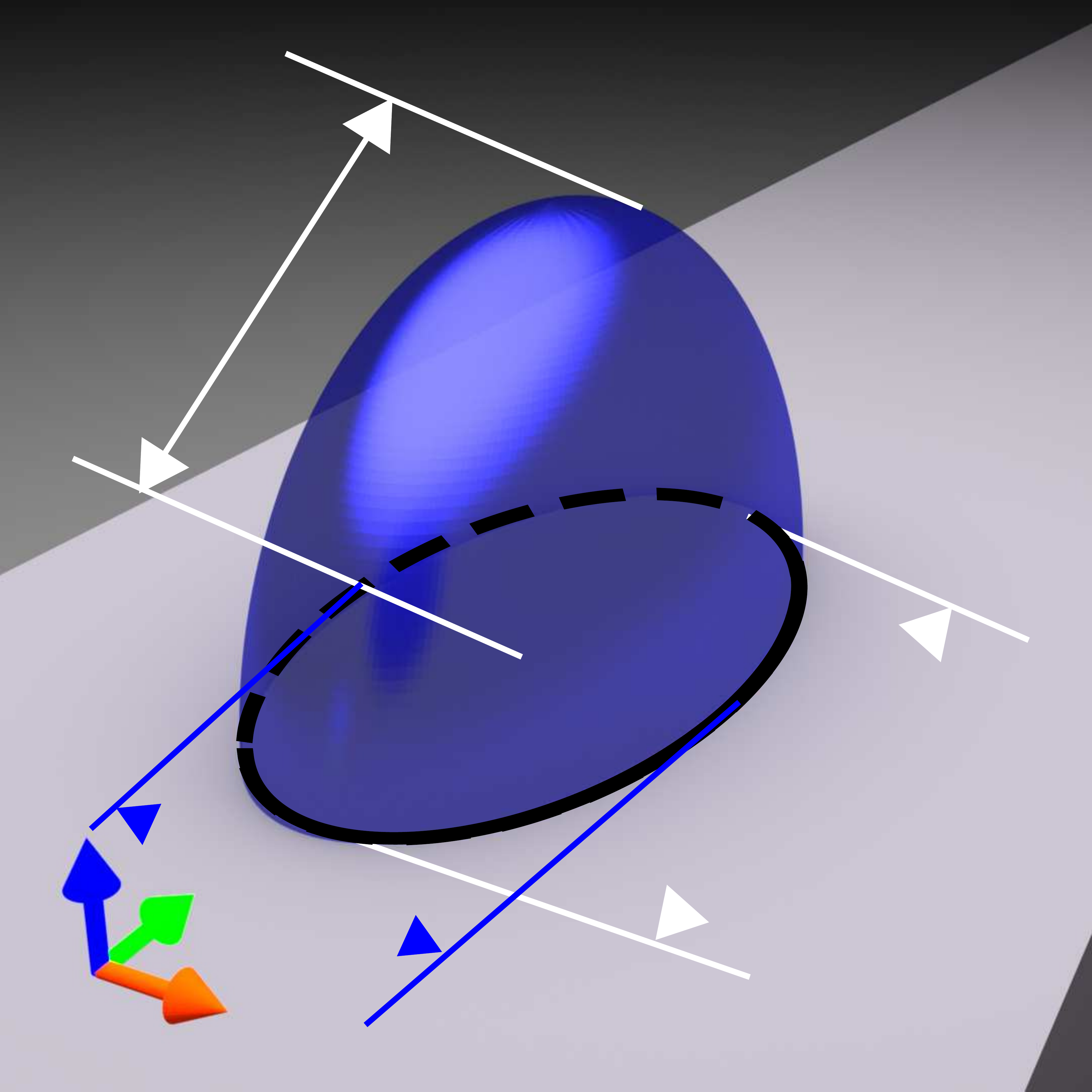}
       \put(0,85){ \color{white} (e) }
         \put(66,25){\rotatebox{-25}{ \small \color{white} 120 }}
         \put(10,75){\rotatebox{-25}{ \small \color{white} 60 }}
         \put(18,18){\rotatebox{-25}{ \small \color{blue} 60 }}
       \end{overpic}
\caption{The nano-antenna shapes and dimensions (all units in nm) considered for this paper (background not to scale).
}
\label{fig:geometries}
\end{figure}

We compare the field focusing effect and calculate optical potentials produced by hemispherical, cubic, conical, toroidal, and hemielliptical nano-antenna geometries, as shown in Fig.~(\ref{fig:geometries}).
Using the field focused by the nano-antenna, we show that a global minimum of the optical potential is attainable, which corresponds to the presence of a stable atom trap.
Unlike the usual method of atomic trapping using focused lasers, this structure can easily be integrated into a miniaturized set-up, and unlike nanoplasmonic lattices for atoms, the all-dielectric system proposed here should reduce the damaging Johnson noise.

{
}

{Fiber-based atom traps are created by illuminating a glass fiber from both end-facets to form a standing wave that, in turn, generates a trapping potential\cite{sorensenPRL2016, corzoPRL2016,vetschPRL2010,gobanPRL2012}, but this type of atom trap is limited to trapping atoms along a line.
The configuration studied here also utilizes a standing wave and, as we discuss below, also enables the possibility of scalable traps in two dimensions.
The trapping fields used in the fiber-based systems use two laser wavelengths, one which is red-detuned with respect to the atomic optical transition ($\lambda_{red} =$~1094~nm) to create an attractive light force, and the other being blue-detuned ($\lambda_{blue} =$~652~nm) to create a repulsive force.
This is referred to as a bichromatic trapping system.
The two fields together form an atom trap at the point where these forces are equal.}

For linearly polarized incident light, each of these potentials is given by~\cite{grimmAIAMaOP2000,mildnerJPBAMOP2018}
\begin{equation}
    \label{eq:opticalpotential}
    U_{{\rm opt},j} = \dfrac{\pi c^2\Gamma}{2 \omega_0^3} \left(\frac{1}{\omega_j-\omega_1}+\frac{2}{\omega_j-\omega_2}\right) I_{{\rm opt}, j}, 
\end{equation}
where $I_{{\rm opt}, j}$ and $\omega_j$ are, respectively, the light intensity at the surface and frequency corresponding to the two light fields ($j$ designating 'red' or 'blue'), and $\Gamma=2\pi\times 5.2$~MHz and $\omega_0$, respectively, are the natural linewidth and average transition frequency of the excited atomic levels in the D line of cesium.
The transition frequencies are $\omega_1$ and $\omega_2$, where the subscript refers to the sublevels $D_1$ ($6^2S_{1/2}\to 6^2P_{1/2}$ with transition wavelength of 852.3\,nm) and $D_2$ ($6^2S_{1/2}\to 6^2P_{3/2}$ with transition wavelength 894.6\,nm)~\cite{steck2003}.
%
%
Calculating the bichromatic optical potential on an ordinary featureless ridge waveguide yields a series of minima spaced by half the optical wavelength along the propagation direction, which is similar to results obtained from related work on atomic trapping using optical fibers\cite{sorensenPRL2016, corzoPRL2016,vetschPRL2010,gobanPRL2012}.

{A sharp optical intensity gradient is required to generate strong localized forces.}
To produce this intensity gradient, we are introducing a perturbation in the boundary conditions, which, in this case, is the nano-antenna.
We use silicon, a material with a high refractive index for VNIR (visible-near infrared) frequencies, in order to concentrate the energy from the waveguide at the near-field of the nano-antenna.
In our system, the evanescent field's need to be more pronounced along the $z$ axis (i.e., towards the cladding where an atom to be trapped), as opposed to the sides (i.e., along $y$), hence we use the fundamental TM mode for the input field\cite{katiyiJLT2017}.
As shown in Eq.~(\ref{eq:opticalpotential}), the optical potential is directly proportional to the field intensity, hence the optical potential can be written as\cite{mildnerJPBAMOP2018,stehleNP2011}
\begin{equation}
U _ { j }(z) = U _ { 0 j } \exp ( - 2  { z } / { z _ { 0 j } } ) ,
\end{equation}
where $U _ { 0j }$ is the potential strength at the surface and $z _ { 0j }$ is the decay length of the evanescent wave.
The potential dependence along $x$ and $y$ are calculated numerically.
In order to take advantage of the evanescent wave's strong intensity gradient, the trap is designed such that the potential minimum is close to the surface.

\begin{figure}[h]
    \centering
    \includegraphics[width=0.55\columnwidth]{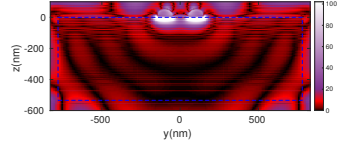}
    \caption{Percent difference between the electric field magnitude in the waveguide without the nano-antenna (|$\mathbf E_{wg}$|) and with the nano-antenna ($|\mathbf E_{na}|$), calculated using $| (\mathbf E_{na}-\mathbf E_{wg})/\mathbf E_{na}|$. The dashed white rectangle indicates the boundaries of the guiding layer.}
    \label{fig:WGNA_difference}
\end{figure}

{We now consider an atomic trapping concept using nano-antennae on a planar waveguide, as the geometry of the waveguide would naturally allow multiple (scalable) trapping potentials either along its length (as a line of atoms) or in two dimensions (as an array of atoms) by adding more nano-antennae.
Placing nano-antennae on top of the waveguide do not affect the guided modes.
This conclusion is supported by the calculating the difference between the electric field magnitude with the nano-antenna ($|\mathbf E_{na}|$) and without (|$\mathbf E_{wg}$|), as shown in Fig.~(\ref{fig:WGNA_difference}).
This illustrates that the guiding layer is hardly perturbed by the presence of the nano-antenna; as can be seen, the maximum perturbation in the waveguide core, excluding the region near the nano-antenna is less than 10\%.
It is well-known that atomic lattices can be used as simulators to study strongly correlated quantum many-body systems\cite{shersonN2010}.
Placing several nano-antennae in an ordered array is important to explore the atom-atom interaction and atom-photon interaction in more than one dimension.}

{The silicon nitride (SiN) ridge waveguide is placed on top of a silica (SiO$_2$) substrate with dimensions of 1.56~$\mu$m along $y$ and 0.534~$\mu$m along $z$.
These dimensions are chosen to ensure that the waveguide has at least one confined quasi-TM mode.
Waveguides with larger cross-sections were not considered because the higher-order modes will add complications coming from multimode interference effects.
The specific dimensions chosen are designed to ensure that the evanescent tail of the fundamental TM mode is sufficiently long-range.
The fields were calculated using Lumerical FDTD\cite{zotero-1756}, with the fundamental TM mode launched into the waveguide.}


{Sub-micron silicon nanostructures have been extensively studied in the field of all-dielectric resonators such as nanocubes, nanocones and other shapes. 
In general, laser-assisted methods have been shown to be effective in nanoparticle fabrication with diameters less than 100~nm\cite{baranovO2017,zywietzNC2014}.
These structures are specifically investigated as alternatives to plasmonic nanoparticles in the manipulation of light-matter interactions, or for applications toward anti-reflection coatings, etc.
}

\begin{figure*}
\centering
\begin{overpic}[width=0.2975\textwidth,tics=20]{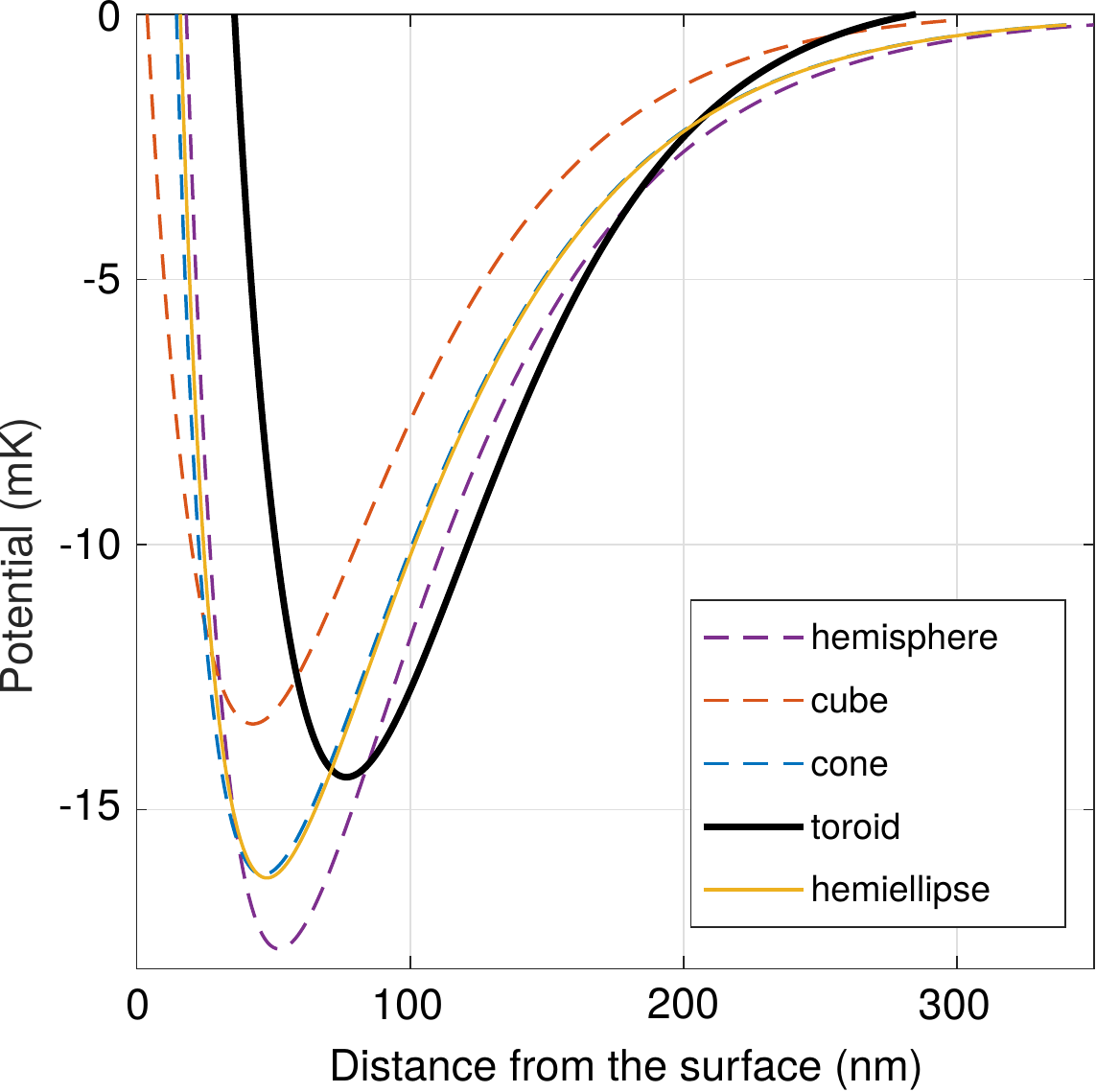}
\put(15,99){ \textbf{(a)} }
\end{overpic}
\hspace{1em}
\begin{overpic}[width=0.3\textwidth]{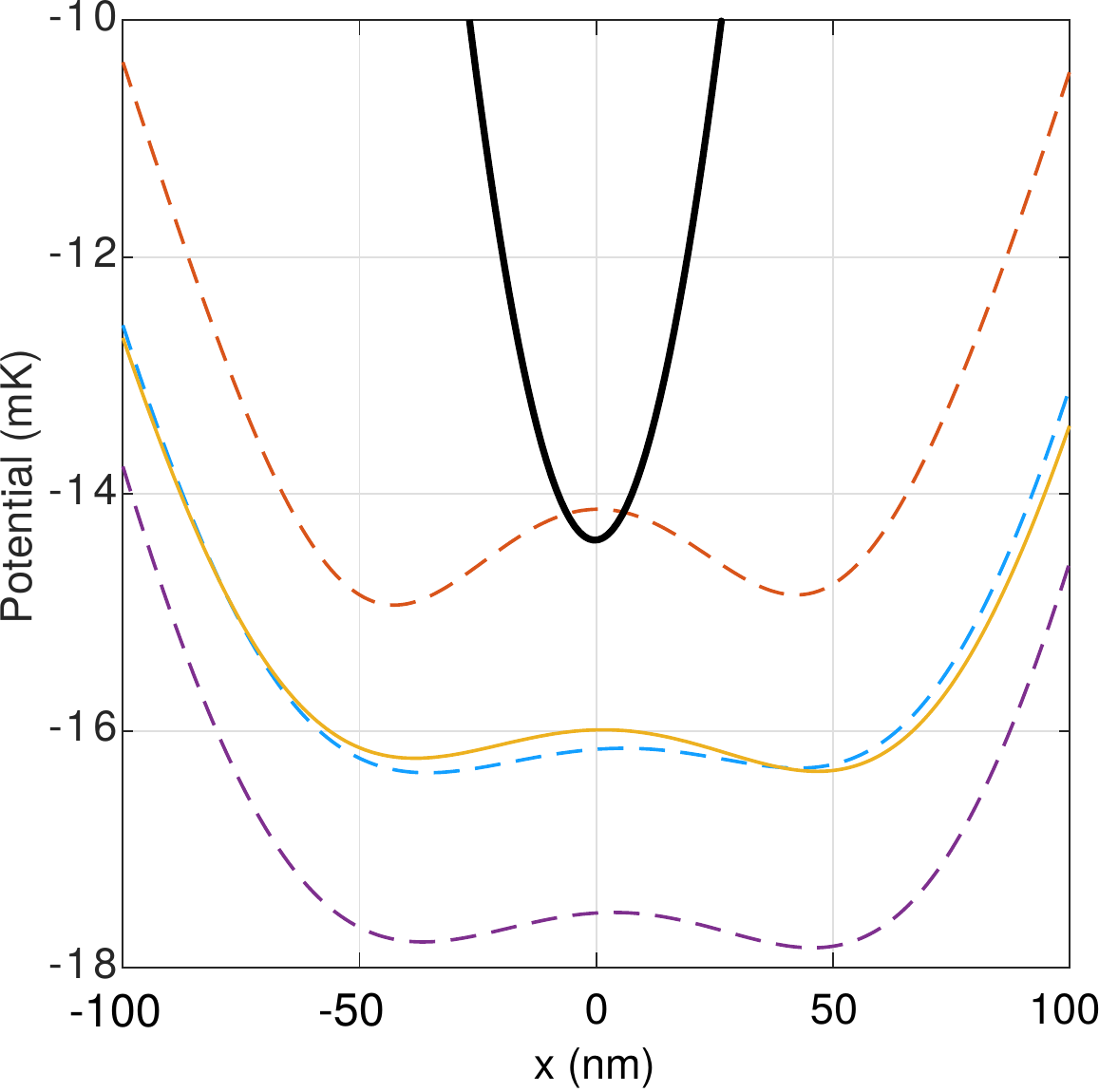}
\put(15,98){ \textbf{(b)} }
\end{overpic}
\hspace{1em}
\begin{overpic}[width=0.3\textwidth]{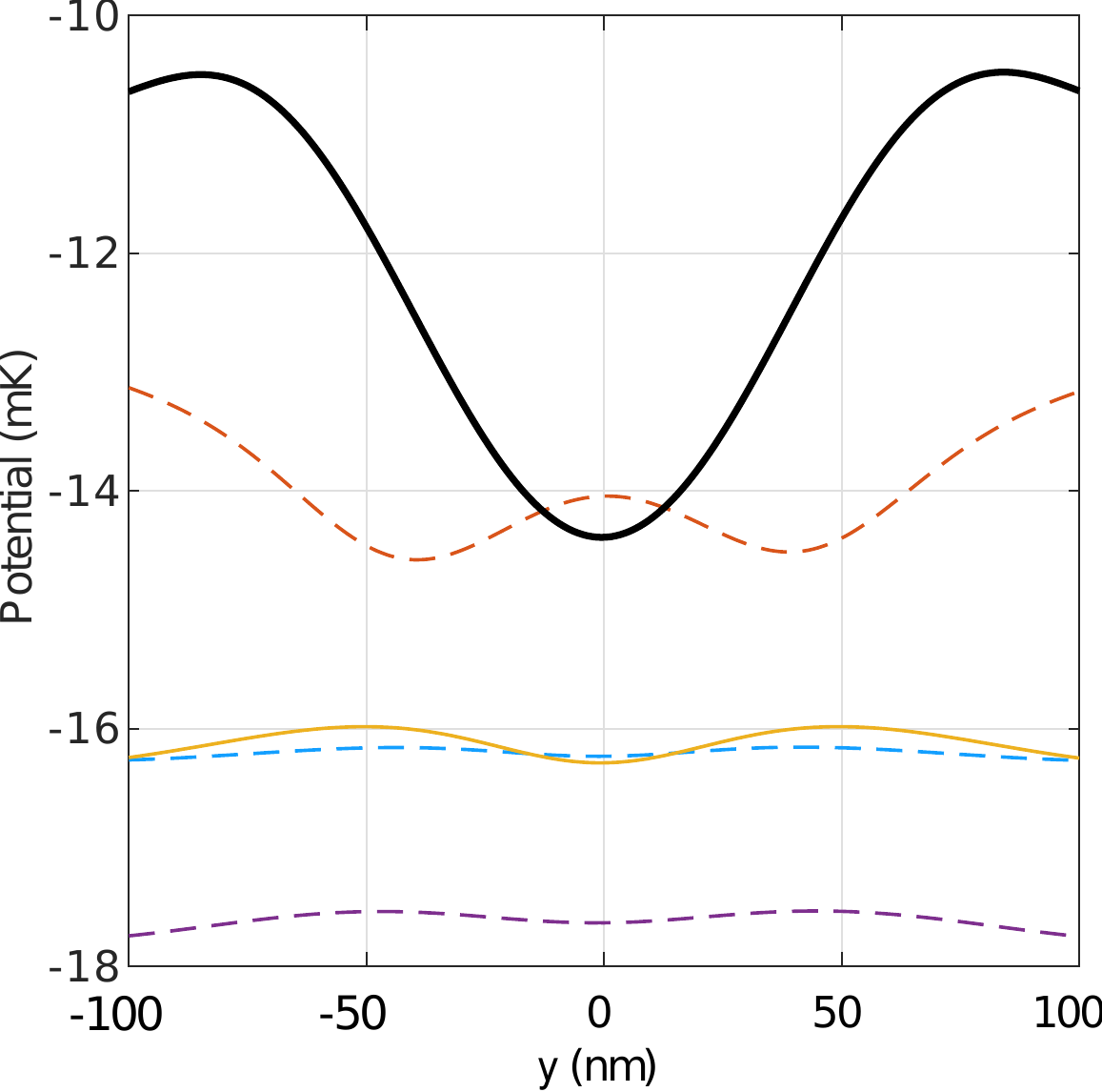}
\put(15,98){ \textbf{(c)} }
\end{overpic}
\caption{(a)~Potential dependence on $z$ where $x=y=0$, (b)~potential dependence on $x$ where $z = z_d$ and $y=0$, and (c)~potential dependence on $y$ where $z = z_d$ and $x=0$.
While these potential cuts are useful for characterizing the potential minimum, the global potential minimum is better presented as a three-dimensional isosurface at a given energy (see Fig.~\ref{fig:isosurface}).
}
\label{fig:dual_input_compilation}
\end{figure*}

We considered a cube and hemisphere for their simplicity, as they can be easily fabricated.
A toroidal structure is also considered for the same reason.
In order to study characteristics that generate better localized fields, we also considered two additional structures: a cone and a hemiellipse.
The cone is considered to see if its sharp peak provides better field localization, and the hemiellipse is considered to see if breaking the symmetry improves the field localization.
The total optical potential is shown along each cartesian axis in Fig.~(\ref{fig:dual_input_compilation}) for each of these shapes.
{
The sizes of the nano-antennae were selected to focus the fields on a subwavelength scale. 
We found that the sizes smaller than about $\lambda/6$ does not localize the fields sufficiently.}

\begin{table*}[h]
\centering
\caption{Numerical results obtained in the simulations. In this table, $P_{red}$ and $P_{blue}$ are the respective powers (in watts) of the field source for the red- and blue-detuned fields, $z_d$ is the distance (in nm) from the potential minimum to the nano-antenna surface, $U_0$ indicates the optical potential minimum (in mK), $U_{gs}$ is the zero-point energy (in mK), and $U_{\text{surf}}$ is the estimated atom-surface potential (in mK).}
\begin{tabular}{rcccccccc}
\hline
\textbf{Structure} & \textbf{$P_{red}$} & \textbf{$P_{blue}$} & \textbf{$z_d$} & \textbf{$U_{0,y}$} & \textbf{$U_{gs,z}$} & \textbf{$U_{\text{surf}}$} \\ \hline
\textbf{hemisphere}                   & 6.1       & 16.9       & 52 & 0.25    & 0.86      & 0.2  \\
\textbf{cube}                         & 6.1       & 16.2       & 43 & 0.53    & 1.07      & 0.4  \\
\textbf{cone}                         & 6.1       & 17.8       & 42 & 0.21    & 0.86      & 0.5  \\
\textbf{toroid}                       & 6.1       & 30.1       & 77 & 3.80    & 1.50      & 0.1  \\
\textbf{hemiellipse}                & 6.1       & 15.6       & 36 & 1.13    & 1.36      & 0.7  \\
\end{tabular}
\label{table}
\end{table*}

More specific data from the simulations are shown in Table~(\ref{table}).
We propose to illuminate the waveguide using the power scaling technique widely used in an integrated photonics:
rare-earth dopants can be used on a segment of the waveguide to amplify the source in order to achieve the listed fields\cite{smithAPB2017}.
{We can also disregard effects due to nonlinearity; the contribution of the nonlinear coefficient\cite{ikedaOEO2008} reads as $Power \times 2.88 * 10^{-7}$, which is negligible compared to the total refractive index.
}

We consider two requirements that must both be met before we can conclude that the potentials may be able to trap cesium atoms.
Firstly, the optical trapping potential near its minimum must not be significantly perturbed by attractive atom-surface potentials that can destroy the trap near the surface and secondly, that the ground-state energy lies within the optical potential well.
The atom-surface potential $U_{\text{surf}}$ is estimated using a Matlab code that calculates approximate Casimir-Polder and van der Waals potentials using the equations described in Sec~III.B of Ref.~\cite{rosenblitPRA2006}, evaluated at $z = z_d$.

The table compares $U_{\text{surf}}$ to $U_{0,y}$ for each nano-antenna shape, where $U_{0,y}$ is the optical trapping potential along the $y$ axis, i.e., the difference between the minimum and maximum of the curves shown in Fig.~(\ref{fig:dual_input_compilation})c.
Noting that the optical trapping force is dependent on the potential gradient, we see that this axis has the weakest such trapping potential because of the comparatively shorter optical wavelength relative to the width of the single-mode waveguide (see Supplementary Figures, page 2, top and bottom rows for the $x$ and $y$ axes, respectively). 
The table also shows the ground-state energy $U_{gs,z}$ calculated assuming a harmonic potential along the $z$ axis (here we use the $z$ axis because it has the highest frequency $\omega_z>\omega_x>\omega_y$ and therefore gives the greatest contribution to the approximate 3D ground-state energy $U_{gs} = \hbar \left(\omega_x + \omega_y+ \omega_z \right) /2$).

\begin{figure}[h]
\centering
\includegraphics[width=0.55\columnwidth]{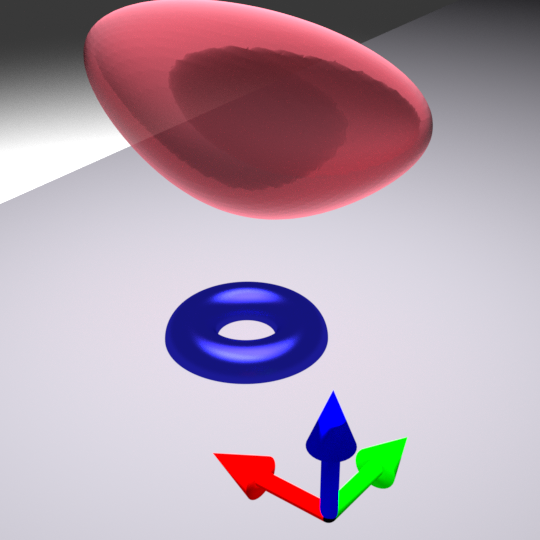}
\caption{Potential isosurface at the zero-point energy for the toroid nano-antenna. The isosurfaces for the remaining geometries are shown in Sec~5. of the Supplementary Materials.
}
\label{fig:isosurface}
\end{figure}

It is apparent from Table~(\ref{table}) that only the toroidal nano-antenna generates a sufficiently strong potential along the $y$ axis to support any bound states, i.e., the ground-state energy $U_{gs,z}$ exceeds $U_{0,y}$ for all the other shapes so they cannot support even a single bound state.
{The uniquely successful toroidal shape was chosen as one of the nano-antenna geometries considered in our study based on a previous paper\cite{xuOE2019} that reports a similar nano-antenna geometry generating an equilibrium force at the midpoint. We also note that the cross-section of the field generated by the toroidal nano-antenna, shown in Figs.~S10~and~S11 of the supplementary materials, is qualitatively similar to that shown in Ref. \cite{xuOE2019}.
} 
The toroidal nano-antenna also exhibits the lowest atom-surface potential $U_{\text{surf}}$ for any of the shapes considered, partially because we have deliberately increased the blue-detuned laser power to push the potential minimum a little further from the surface; as it is only a small fraction of the potential minimum at these distances, we do not consider $U_{\text{surf}}$ further.
In Fig.~(\ref{fig:isosurface}), we show an isopotential surface at the ground-state energy $U_{gs,z}$ for the toroidal nano-antenna.
This isopotential surface is completely closed, indicating at least qualitatively that the cesium atom could indeed be trapped by the optical potential generated near such a nano-antenna.
This more demanding test again eliminates any of the other geometries considered, and we conclude that the highly symmetric toroidal shape is the most suitable for atom trapping.
{Breaking the symmetry, in the case of the hemiellipse, improves the trap parameters, but not sufficiently for generating a closed isosurface at the zero-point energy.
}

For future perspective, bringing the atoms to distances below 100\,nm from the nano-antenna surface may allow exploration of Casimir-Polder and van-der-Waals forces for the unique shapes considered here.
This is a challenging theoretical problem as well as an experimental one, since the surface shape affects these forces qualitatively as well as quantitatively but is outside the scope of the current paper.
It would also be interesting to explore heating effects that may be generated in such systems, as well as other possible mechanisms that can cause atoms to leave the trap.

{The temperature-associated effects are the subject of our future work, also see the supplementary material. We would emphasize, however, that the potential trap generated on top of our all-dielectric system is essentially generated by the evanescent fields of the waveguide. For such a waveguide, the power carried by the evanescent field is about 0.08 of the overall power launched into the waveguide\cite{katiyiJLT2017}, hence the cladding region of the waveguide does not experience significant heating.}


In conclusion, we have presented an optical trapping concept in an all-dielectric system, where a silicon nano-antenna produces a trapping potential for atoms in  a chip-scale configuration.
We numerically explored the dependence of the bichromatically constructed trapping potential on the nano-antenna shape.
Our results show that the near-field induced potential landscape can be tailored to sub-wavelength dimensions by  manipulating the evanescent field using different nano-antenna geometries on top of the waveguide.
We found that a toroid generates an all-optical Cs atom trap directly above the nano-antenna.
This trapping set-up can be used as a method for trapping an array of single atoms with the advantage of easy integration within a silicon photonic platform, thereby paving the way toward fascinating atom-photon applications on a chip.

\section*{Acknowledgements}
This work has been supported by the Israeli Innovation Authority, Grant. No. 69073.
A.S.S acknowledges the support of the Russian Fund for Basic Research within the projects 18-02-00414, 18-52-00005. Simulations of the optical trapping has been supported by the Russian Science Foundation (Project No. 18-72-10127).
A.S.A. thanks Mark Keil, Yonathan Japha, and David Groswasser for fruitful discussions.

\vspace{1em}

\textbf{Disclosures}: The authors declare no conflicts of interest.


\end{document}